\documentclass{llncs}
\usepackage{makeidx}  % allows for indexgeneration
\renewcommand{\deg}{\(\mathsurround=0pt{}^\circ\)}
\usepackage[utf8]{inputenc}
\usepackage{float} %for fixing a figure position in the text

\usepackage[table,xcdraw]{xcolor}
\usepackage{subcaption} 

\newcolumntype{s}{>{\columncolor[HTML]{AAACED}} p{1cm}}
 
\arrayrulecolor[HTML]{DB5800}

\usepackage{amssymb}
\usepackage{graphicx}
\usepackage{rotating}
\usepackage{listings}
\usepackage{textcomp}
\usepackage{color}
\usepackage{tikz}
\usetikzlibrary{arrows,automata}
\definecolor{codegreen}{rgb}{0,0.6,0}
\definecolor{codegray}{rgb}{0.5,0.5,0.5}
\definecolor{codepurple}{rgb}{0.58,0,0.82}
\definecolor{backcolour}{rgb}{0.95,0.95,0.92}
\definecolor{deepblue}{rgb}{0,0,0.5}
\definecolor{deepred}{rgb}{0.6,0,0}
\definecolor{deepgreen}{rgb}{0,0.5,0}
 
\lstdefinestyle{mystyle}{
    backgroundcolor=\color{backcolour},   
    commentstyle=\color{codegreen},
    keywordstyle=\color{magenta},
    numberstyle=\tiny\color{codegray},
    stringstyle=\color{codepurple},
    basicstyle=\footnotesize,
    breakatwhitespace=false,         
    breaklines=true,                 
    captionpos=b,                    
    keepspaces=true,                 
    numbers=left,                    
    numbersep=5pt,                  
    showspaces=false,                
    showstringspaces=false,
    showtabs=false,                  
    tabsize=2
}
 
\DeclareFixedFont{\ttb}{T1}{txtt}{bx}{n}{9} % for bold
\DeclareFixedFont{\ttm}{T1}{txtt}{m}{n}{9}  % for normal
\newcommand\pythonstyle{\lstset{
language=Python,
basicstyle=\ttm,
otherkeywords={self},             % Add keywords here
keywordstyle=\ttb\color{deepblue},
emph={MyClass,__init__},          % Custom highlighting
emphstyle=\ttb\color{deepred},    % Custom highlighting style
stringstyle=\color{deepgreen},
frame=tb,                         % Any extra options here
showstringspaces=false            % 
}}

\lstnewenvironment{python}[1][]
{
\pythonstyle
\lstset{#1}
}
{}

\newcommand\pythoninline[1]{{\pythonstyle\lstinline!#1!}}
\usepackage[colorlinks = true,
            linkcolor = blue,
            urlcolor  = blue,
            citecolor = blue,
            anchorcolor = blue]{hyperref}

\usepackage{url}
\urldef{\mailsa}\path|{irina.rychkova, fabrice.boissier}@univ-paris1.fr |

\newcommand{\keywords}[1]{\par\addvspace\baselineskip
\noindent\keywordname\enspace\ignorespaces#1}

\graphicspath{{Figures/}}

\begin{document}

\mainmatter  % start of an individual contribution

\title{A Pragmatic Approach for Measuring Maintainability of DPRA Models}

\titlerunning{Maintainability of DPRA Models}

\author{Irina Rychkova\inst{1}\and Fabrice Boissier\inst{1} \and Hassane Chraibi\inst{2} \and Valentin Rychkov\inst{2}}
\authorrunning{Irina Rychkova  \and Fabrice Boissier \and Hassane Chraibi \and Valentin Rychkov}

\institute{Universit\'e Paris 1 Panth\'eon-Sorbonne,\\
12, Place du Panth\'eon, 75005 Paris, France
\and
EDF R\&D, EDF Lab Paris-Saclay, \\
7 Bd. Gaspart Monge, Palaiseau, France
}

\toctitle{Maintainability of DPRA models}
\tocauthor{Rychkova et al}
\maketitle

\begin{abstract}
Dynamic Probabilistic Risk Assessment (DPRA) is a powerful concept that is used to evaluate design and safety of complex industrial systems. A DPRA model uses a conceptual system representation as a formal basis for simulation and analysis. In this paper we consider an adaptive maintenance of DPRA models that consist in modifying and extending a simplified model to a real-size DPRA model. We propose an approach for quantitative maintainability assessment of DPRA models created with an industrial modeling tool called PyCATSHOO.
We review and adopt some metrics from conceptual modeling, software engineering and OO design for assessing maintainability of PyCATSHOO models. On the example of well-known "Heated Room" test case, we illustrate how the selected metrics can serve as early indicators of model modifiability and complexity. These indicators would allow experts to make better decisions early in the DPRA model development life cycle.

\keywords{maintainability metrics, conceptual models, object oriented design, DPRA models}
\end{abstract}

\section{Introduction}
\label{intro}
%1. Intro
%Design of a complex industrial system with respect to its safety requirements is an important endeavor for both engineers and R\&D. 
Dynamic Probabilistic Risk Assessment is a powerful concept that is used to evaluate design and safety of complex systems where the static reliability methods like fault trees find their limits \cite{aldemir2013survey}. A DPRA model combines the characteristics of conceptual model and a software application: it formally describes some aspects of the physical world (for example, a complex mechanical system) for  purposes of understanding and communication \cite{mylopoulos1992conceptual}; it serves a formal basis for further system simulation and analysis. 
For feasibility, proof of concept, algorithm benchmarking and other preliminary studies simplified DPRA models are used. Adaptive maintenance is an important part of a DPRA model life cycle: it consists in modifying and extending a simplified model to a real-size DPRA model.
%Architecture and conceptual modeling principles are usually out of concern for simplified DPRA models. 
%Whereas a design of a simplified DPRA model for a "small cut" of the system can be done relatively easily, scaling for a real-size DPRA problem is challenging due to the scope and complexity of real system interactions. 

In this work, we propose an approach for maintainability assessment for DPRA models created with an industrial modeling tool called PyCATSHOO.
We define a set of metrics that can serve as early indicators of PyCATSHOO model modifiability and complexity.

PyCATSHOO is a tool dedicated for dependability analysis of hybrid systems developed and used by EDF \cite{chraibi2013,chraibi2016}. 
%Hybrid systems are  systems where deterministic continuous phenomena and stochastic discrete behaviour are equally important \cite{RychkovKawahara}
PyCATSHOO models are executable modules that can be written in Python or C++ and interpreted by PyCATSHOO engine. 

We review some well-known metrics from conceptual modeling, software engineering and object-oriented design, including size measures (LOC), complexity measures (McCabe's Cyclomatic Complexity \cite{mccabe1976complexity}), lexical measures (Halstead \cite{halstead1977elements}), Maintainability Index (MI) \cite{oman1992metrics} and OO-specific measures \cite{chidamber1994metrics}. Based on our review, we select and adapt a set of metrics applicable to DPRA models and PyCATSHOO models in particular. 

We make an assumption that \textit{the selected metrics can show a difference between two PyCATSHOO model designs already at the early model development life cycle},  helping experts to make better decisions. 

To validate this assumption, we apply selected metrics to assess and compare two designs of the Heated Room system. Heated Room is a well-known test case reported in \cite{bouissou2013critical,bouissou2007comparison}. It describes a system that consists of a room and a heater that can switch on and off to maintain the ambient temperature. This example illustrates a hybrid system that combines deterministic continuous phenomena (i.e., temperature evolution) and stochastic discrete behaviour (i.e., functioning of a heater). 

We create two sets of PyCATSHOO models as illustrated in Table\ref{models}: 
Set 1 follows the original design ideas from \cite{chraibi2013}, Set 2 represents an alternative model design promoting the low coupling design principle. 
The corresponding models in two sets are semantically equivalent, i.e., they demonstrate the same simulation traces. We compare measurements for the model sets and discuss the results.

The reminder of this paper is organised as follows: Section 2 presents DPRA models and PyCATSHOO modeling tool; Section 3 discusses the related works on maintainability assessment; Section 4 presents our approach for  maintainability assessment of PyCATSHOO models; In Section 5 we describe  two alternative designs of the PyCATSHOO models for the Heated Room test case. We assess maintainability of these designs and discuss our results in Section 6; Section 7 presents our conclusions.
\vspace*{-\baselineskip}
\begin{table}[]
\centering
\caption{PyCATSHOO models of the Heated Room test case}
\label{models}
\begin{tabular}{|l|l|l|}
\hline
\rowcolor[HTML]{EFEFEF} 
\multicolumn{1}{|c|}{\cellcolor[HTML]{EFEFEF}\textbf{Model description:}} & \multicolumn{1}{c|}{\cellcolor[HTML]{EFEFEF}\textbf{\begin{tabular}[c]{@{}c@{}}Set 1 \\ (Design 1)\end{tabular}}} & \multicolumn{1}{c|}{\cellcolor[HTML]{EFEFEF}\textbf{\begin{tabular}[c]{@{}c@{}}Set 2 \\ (Design 2)\end{tabular}}} \\ \hline
Initial model: 1 heater + 1 room & Case 0 & Case 0a \\ \hline
System level modificaiton: 4 independent heaters H0..H3 + 1 room & Case 1 & Case 1a \\ \hline
\begin{tabular}[c]{@{}l@{}}Component level modification: standby redundancy of heaters: \\ H0 - main; H1..H3 - backups\end{tabular} & Case 2 & Case 2a \\ \hline
\end{tabular}
\end{table}
%\vspace*{-8mm}
 %Our finding demonstrate that the selected metrics show the difference between original and alternative model designs and can provide an expert with an insight about model complexity and modifiability early at DPRA model development phase.

%On the example of a PyCATSHOO model of Heated Room, we illustrate how the selected metrics can be used to compare different DPRA model designs.

%On the example of a DPRA model for Heated Room, we study how the  maintainability and complexity vary depending on the architectural choices made early in the modeling. 
%We compare a PyCATSHOO model of Heated room created without considering any particular design principles as reported in \cite{chraibi2013} and its variants where we implement the low coupling design principle using some well-known patterns of object-oriented design (i.e., Mediator, Observer, Chain of responsibility \cite{wolfgang1994design}).

%We show that the selected metrics can serve as early indicators of model modifiability and complexity. These indicators would allow experts to make better decisions early in the DPRA model development life cycle.  

%Our results show that the "right" architectural decisions made for a simplified DPRA model pay off, improving scalability and reducing complexity of the DPRA model. 
%\section{Background}
\label{background}
%2. Background
\section{DPRA Models and PyCATSHOO Modeling Tool}
\label{pra}
%  2.1 DPRA Models
%Dynamic Probabilistic Risk Assessment (DPRA) is a powerful concept that is used to evaluate design and safety of complex systems where the static reliability methods find their limits \cite{aldemir2013survey}. 
  %2.2 PyCATSHOO models
%The safety requirement of its nuclear and hydraulic fleet, has allowed 
EDF has a long-standing experience in using and developing DPRA tools for complex systems. PyCATSHOO is one of the tools developed over last few decades at EDF R\&D. PyCATSHOO implements the concept of Piecewise Deterministic Markov Process (PDMP) using distributed stochastic hybrid automata. The principles of this paradigm are described in details in \cite{chraibi2013}. 

PyCATSHOO models are used for advanced risk assessment of EDF's hydro and nuclear electrical generation fleet. 
PyCATSHOO is grounded on the Object-Oriented (OO) and Multi-Agent System (MAS) paradigms\cite{michel2009multi}. %According to \cite{michel2009multi}, MAS is a paradigm which aims to comprehend the overall system behavior through the interaction between its constituent entities that evolve autonomously and act proactively.
Following the OO paradigm, PyCATSHOO defines a system, its subsystem or component as a class - an abstract entity that can be instanced into objects. The latter are concrete entities which communicate by message
passing and that are able to perform actions on their own encapsulated states. This paradigm has been successfully implemented for modeling and analysis of stochastic hybrid systems as reported in \cite{meseguer2006specification}.
%This paradigm is useful to deal with systems complexity in the large scale software development area. It has been also successfully implemented for modeling and analysis of stochastic hybrid systems as reported in \cite{meseguer2006specification}.
Following MAS, PyCATSHOO models a system as a collection of objects
with a reactive agent-like behavior. A reactive agent acts using a stimulus-response mode where only its current state and its perception of its environment are relevant. %A reactive agent doesn’t take history into account. It has therefore a Markovian behavior.
\subsection{DPRA Modeling with PyCATSHOO}
\label{concept}    
  %2.4 Conceprual or Software?
PyCATSHOO offers a flexible modeling framework that allows for defining generic components (classes) of hybrid stochastic automata to model a given system or a class of systems with a particular behaviour. 
A hybrid stochastic automaton may exhibit random transitions between its states according to a predefined probability law. It may also exhibit deterministic transitions governed by the evolution of physical parameters.

A modeling process with PyCATSHOO can be summarised as follows: 
\begin{itemize}
\item {Conceptual level:} A system is decomposed into elementary subsystems, components.
\item {Component level:} Each system component is described with a set of hybrid stochastic automata, state variables and message boxes. Message boxes ensure message exchanges between components. 
\item {System level:} To define the system, the components are instantiated from their corresponding classes. Component message boxes are connected according to the system topology. 
\end{itemize}
A DPRA model in PyCATSHOO combines the characteristics of conceptual model and a software application: it formally describes some aspects of the physical world (for example, a complex mechanical system) for  purposes of understanding and communication \cite{mylopoulos1992conceptual}; it serves a formal basis for further system simulation and analysis. 

%Today, 
PyCATSHOO offers Application Programming Interfaces (APIs) in Python and C++ languages. %PyCATSHOO models are expressed in Python language has no explicit graphical notation. 
Once the model is designed, the system behaviour is simulated. An analyst needs to use Monte Carlo sampling if the system exhibits random transitions. Sequences (time histories of the system evolution) that lead to desirable end states are traced and clustered.
%\subsection{Design and Maintainability}
%\label{design}  

In \cite{aldemir2013survey}, various modeling tools for DPRA are discussed. 
Whereas some modeling tools propose a visual modeling interface, model complexity and high development and maintenance costs are considered the main obstacle for efficient use of DPRA models in industry \cite{DPRA_Difficulty}. Quantitative measures of model maintainability would be of a great value, helping the experts to make better decisions early in the DPRA model development life cycle. %A tool for reasoning about model quality and guidelines for engineers for improving such quality would be of a great value.

%For feasibility, proof of concept, algorithm benchmarking and other preliminary studies simplified DPRA models are used. Reduction of number of system components, number of attributes that characterise a component, number and types of dependencies between components are some examples of simplification. Quality and design  of the underlying conceptual model is usually out of concern for simplified DPRA models. 

%Whereas a design of a simplified DPRA model for a "small cut" of the system can be done relatively easily, scaling for a real-size DPRA problem is challenging due to the scope and complexity of real system interactions.

%Maintainability of DPRA models and PyCATSHOO models in particular remains challenging. 
  %2.3 Design and Maintanability

\section{Maintainability of Models: State of the Art}
\label{state}
%3. Maintanability of models
ISO 9000 is a set of international standards on quality management. It defines quality as \textit{"the totality of features and characteristics of a product or service that bear on its ability to satisfy stated or implied needs"}  \cite{hoyle2001iso}. 
Maintainability is a quality characteristic that is defined as \textit{"a set of attributes that bear on the effort needed to make specified modifications."}  
%in engineering, maintainability is closely associated with bug fixing, but should be better called the ‘change ability’. \cite{gilb2008designing}. 
\subsection{ Maintainability in Software Engineering}  
In Standard Glossary of Software Engineering Terminology \cite{eee1990standard} software maintainability is defined as “the ease with which a software system or component can be modified to correct faults, improve performance or other attributes, or adapt to a changed environment”.

%ISO/IEC 25010  is an international standard for evaluating quality of software products. This standard defines maintainability as one of the product quality characteristics that consists of five sub-characteristics: Modularity, Reusability, Analyzability, Modifiability and Testability \cite{ISO25010}.

% It defines a quality in use model and a product quality model \cite{ISO25010}. Maintainability is one of the product quality characteristics. It consists of five sub-characteristics: Modularity, Reusability, Analyzability, Modifiability and Testability. 

%Modifiability - degree to which a product or system can be effectively and efficiently modified without introducing defects or degrading existing product quality. In our work, this sub-characteristic is most relevant for DPRA models. It is also related to complexity \cite{li1987empirical}.

According to ISO/IEC 25010, maintainability is a sub-characteristic of product quality that can be associated with more concrete, "measurable", quality metrics. Various types of metrics accepted in SE include: size metrics (e.g., Lines of Code), lexical metrics (e.g., Halstead software science metrics \cite{halstead1977elements}), metrics based on control flow graph (e.g., Mc'Cabe's cyclomatic complexity \cite{mccabe1976complexity}) and others.

Metrics specific to Object-Oriented paradigm focus on OO concepts such as object, class, attribute, inheritance, method and message passing. Chidamber \& Kemerer’s OO metrics \cite{chidamber1994metrics} are among the most successful predictors in SE. They include metrics focused on object coupling. 
In \cite{li1993object} ten software metrics and their impact on the maintainability are studied. The size metrics and McCabe's complexity are placed among reliable indicators of maintainability. In \cite{riaz2009systematic}, a systematic review
of software maintainability prediction models and metrics is presented. According to this review, a list of successful software
maintainability predictors include Halstead metrics, McCabe's complexity and size metrics.

Abreu’s Metrics for Object-Oriented Design (MOOD) are presented in \cite{e1995mood} and evaluated in \cite{harrison1998evaluation}. According to MOOD, various mechanisms like encapsulation, inheritance, coupling and polymorphism can influence reliability or maintainability of software. 

The maintainability index (MI) is a compound metric \cite{oman1992metrics} that helps to determine how easy it will be to maintain a particular body of code. MI uses the Halstead Volume, Cyclomatic Complexity, Total source lines of code.  

The models and metrics above address the maintainability at later phases of software development life cycle. In the next part of this section, we consider maintainability at design phase. In particular, maintainability of conceptual models.
\vspace*{-\baselineskip}
\subsection{Maintainability in Conceptual Modeling}
Conceptual modeling is the activity of formally describing some aspects of the physical and social world around us for the purposes of understanding and communication \cite{mylopoulos1992conceptual}. %The conceptual model plays an important role in the overall system development life cycle. 
Though, in practice, conceptual models are widely used in design and development of IS, they also play an important role in the overall system development life cycle and can be evaluated as representations of the "real world".
We adopt the definition from  \cite{moody2005theoretical}, where conceptual modelling is defined as a design discipline and conceptual models are considered as \textit{"design artifacts used to actively construct the world rather than simply describe it"}.
While ISO/IEC 25010  family standards is widely accepted for evaluating software systems, no equivalent standard for evaluating quality of conceptual models exist.

In \cite{moody2005theoretical,nelson2012conceptual,cherfi2002conceptual} frameworks for conceptual modeling quality are presented. In \cite{nelson2012conceptual}, the empirical quality %is used as a measure of the readability of a conceptual representation \cite{krogstie2006process} 
is considered as a good indicator of the maintainability. In \cite{cherfi2002conceptual} maintainability of conceptual schema is defined as \textit{"the ease with which the conceptual schema can evolve"}. The maintainability of a conceptual schema implies the study of modeling elements cohesion and can be related to the \textit{modifiability} sub-characteristic from ISO/IEC 25010. Quantitative analysis and estimation of conceptual model quality remains challenging due to lack of measurement \cite{moody2005theoretical}. 

%In \cite{nelson2012conceptual}, a conceptual modeling quality framework is presented. This framework is grounded on two well-known quality frameworks: the framework of Lindland, Sindre, and Solvberg \cite{lindland1994understanding} and the framework of Wand and Weber based on Bunge’s ontology \cite{wand1990ontological}. In this framework, the empirical quality is used as a measure of the readability of a conceptual representation \cite{krogstie2006process} is mentioned as a good indicator of the maintainability. 
%However, no concrete measures of empirical quality are recommended.
%In \cite{cherfi2002conceptual} a Framework for Conceptual Quality Evaluation is presented. Maintainability of conceptual schema is defined as \textit{"the ease with which the conceptual schema can evolve"}. The maintainability of a conceptual schema implies the study of modeling elements cohesion and can be related to the \textit{modifiability} sub-characteristic from ISO/IEC 25010. 
%The framework \cite{cherfi2002conceptual} does not propose specific measures for conceptual schema maintainability. 
An important body of knowledge is developed adopting and extending the metrics from software engineering to the conceptual modeling. These metrics are used to estimate quality of conceptual models and UML diagrams in particular. In \cite{genero2005survey}, a survey of metrics for UML class diagrams is presented. In \cite{marchesi1998ooa}, a suite of metrics for UML use case diagrams and complexity of UML class diagrams is proposed. Directly measurable metrics such as number of usecases, actors, classes, subclasses, dependencies e.t.c., 
are used as an early estimate of development efforts, implementation time and cost of the system under development. 
In \cite{genero2000early,genero2007building,genero2002empirical}, 
a set of metrics to measure structural complexity of UML class diagram  are proposed and validated. The authors promote an idea that the structural properties (such as structural complexity and size) of a UML class diagram have an effect on its cognitive complexity. High cognitive complexity reduce model understandability, modifiability and affect its maintainability by the consequence. The authors propose a set of metrics based on OO design state of the art. 
The experimental work demonstrates that the good indicators of class diagram maintainability include: number of associations, aggregations, generalization, total number of aggregation and generalization hierarchies, the maximum value of direct inheritance tree. 
In \cite{genero2002defining} the same group of researchers proposes metrics for measuring complexity of UML statechart diagrams.  The authors show that number of states, transitions and activities metrics are correlated with the understandability.

In \cite{rizvi2010maintainability}, a multivariate linear model ‘Maintainability Estimation Model for Object-Oriented software in Design phase’  (MEMOOD) is presented. This model  estimates the maintainability of class diagrams in terms of their understandability and modifiability. Modifiability is evaluated using the number of classes, generalisations, aggregation hierarchies, generalization hierarchies, direct inheritance tree.  
%Compared to directly measurable properties of conceptual diagrams, the authors of \cite{} propose aesthetic metrics as indicators of class and sequence diagrams maintainability. These metrics cover visual aspects of a diagram such as edge crossing, edge length, diagram layout and relative positioning of elements on the diagram. 

%According to \cite{hoyle2001iso}, maintainability can be associated with a mean time to repair measure. Maintainability prediction can be also based on probability theory and statistical analysis of products (models in our case).
  
%after a thorough review of some of the existing OO measures, applicable to class diagrams at high-level design stage [6],[7],[8][9] we have proposed [10],[11] a set of UML class diagram structural complexity measures based on the use of UML. relationships (associations, generalizations, aggregations and dependencies), see table 1, where also, traditional metrics such as Number of Classes, Number of Methods and Number of Attributes are included. 

%In this paper, we will introduce and analyse a set of an existent object oriented metrics that can be applied for assessing class diagrams complexity at the initial phases of the object oriented development life cycle. We also define our own proposal for new ones. 

%Early Measures for UML Class Diagrams.. Available from: https://www.researchgate.net/publication/220264059_Early_Measures_for_UML_Class_Diagrams [accessed Apr 13, 2017].
\section{Maintainability Assessment of PyCATSHOO Models}

Intrinsically, maintainability is associated with the maintenance process, which represents the majority of the costs of a software development life-cycle \cite{riaz2009systematic}. It is valid for the model development life cycle as well. % Improving design quality by making analysis and assessment can be considered among preventive actions for improving final product quality according to ISO 9000. 

Assessment of conceptual model maintainability can help designers to anticipate model complexity, to incorporate required enhancements and to improve consequently the maintainability of the final software \cite{rizvi2010maintainability}. 
%A DPRA model in PyCATSHOO combines the characteristics of conceptual model and a software application. 
In this work we adapt and  apply  several metrics from SE and OO design to evaluate the maintainability of PyCATSHOO models early at the model development life cycle.

Adapting a maintainability definition from \cite{eee1990standard}, we define maintainability of a DPRA model as \textit{the ease with which a model or its component can be modified to correct faults, improve performance or other attributes, or adapt to a changed environment}.

According to \cite{ISO25010},  maintainability consists of five sub-characteristics: modularity, reusability, analyzability, modifiability and testability.  \textit{Modifiability} sub-characteristic is most relevant for DPRA models; it specifies a degree to which a product (a DPRA model in our case) can be effectively and efficiently modified without introducing defects. 

%Maintainability of a DPRA model can be also associated with its modifiability and understandability.
Similar to software system maintainability, various types of maintainance for DPRA models can be identified: 
\begin{itemize}
\item{Adaptive}  - modifying the model to cope with changes in the  environment;
\item{Perfective} – improving or enhancing a model to improve overall performance; 
\item{Corrective} – diagnosing and fixing errors, possibly ones found by users;
\item{Preventive} – increasing maintainability or reliability to prevent problems in the future (i.e., model architecture, design).
\end{itemize}
In this work, we focus on \textit{adaptive maintenance} that reflects a transformation of simplified DPRA models to real size models.
\subsection{Adaptive Maintainability in PyCATSHOO Models}
Different classes of modifications can be introduced into a  PyCATSHOO model while adapting it to a real size model.
In this work, we consider two classes of PyCATSHOO model modifications:
\begin{enumerate}
    \item Component level modifications - modifications that consist in adapting structure and/or behavior of a model component (e.g., state variables, PDMP equation methods for continuous variables, start/stop conditions for PDMP controller, transition conditions, message boxes etc.).
    \item System level modifications - modifications that consist in adapting structure and topology of the system (e.g., number of component instances, their parameters, dependencies, connections via message boxes etc.).
\end{enumerate}
Each modification class can be related to different requirements and consequently different technical solutions \cite{gilb2008designing}. We argue that the "right" architectural and design decisions made for a simplified DPRA model pay off, improving maintainability and reducing complexity of the real-size DPRA model and PyCATSHOO models in particular. The maintainability metrics can serve as indicators for DPRA domain experts in order to assess their design and architectural decisions early in the modeling. 

\subsection{Selecting Maintainability Metrics}
%after a thorough review of some of the existing OO measures, applicable to class diagrams at high-level design stage [6],[7],[8][9] we have proposed [10],[11] a set of UML class diagram structural complexity measures based on the use of UML. relationships (associations, generalizations, aggregations and dependencies), see table 1, where also, traditional metrics such as Number of Classes, Number of Methods and Number of Attributes are included. 

%In this paper, we will introduce and analyse a set of an existent object oriented metrics that can be applied for assessing class diagrams complexity at the initial phases of the object oriented development life cycle. We also define our own proposal for new ones. 
%Early Measures for UML Class Diagrams.. Available from: https://www.researchgate.net/publication/220264059_Early_Measures_for_UML_Class_Diagrams [accessed Apr 13, 2017].
After a review of some existing metrics focusing on maintainability, we select the set of metrics applicable to DPRA models and PyCATSHOO models in particular. We propose a new metric to measure relative modifications - RLOC. We summarize the selected metrics in Table \ref{metrics}.
%In this set, we include:
%\begin{enumerate}
%    \item The metrics that reflect the OO specifics of PyCATSHOO model
%    \item The metrics associated with modifiability
%    \item The metrics associated with complexity
%\end{enumerate}
Possibility of seamless integration of metrics into current modeling process with PyCATSHOO and availability of measurement tools are  important criteria for our metrics selection. Radon\footnote{\url{http://radon.readthedocs.io/en/latest/intro.html}} is a Python tool which computes various code metrics. Radon supports raw metrics, Cyclomatic Complexity, Halstead metrics and the Maintainability Index.  Cloc\footnote{\url{https://sourceforge.net/projects/cloc/?source=typ_redirect}} - Count Lines of Code - is a tool that counts blank lines, comment lines, and physical lines of source code in many programming languages. Given two versions of a code base, cloc can compute differences in source lines.
%In Table \ref{metrics}, we summarize the metrics relevant for the maintainability assessment of PyCATSHOO models.
\vspace*{-8mm}
\begin{table}[]
\centering
\caption{Metrics for Maintainabiity Assessment of PyCATSHOO models}
\label{metrics}
\begin{tabular}{|l|l|}
\hline
\rowcolor[HTML]{C0C0C0} 
\multicolumn{1}{|c|}{\cellcolor[HTML]{C0C0C0}\textbf{Metrics:}} & \multicolumn{1}{c|}{\cellcolor[HTML]{C0C0C0}\textbf{Tool support:}} \\ \hline
LOC (lines of code), RLOC (relative modifications) & cloc, radon \\ \hline
Cyclomatic Complexity (CC) & radon \\ \hline
Halstead (vocabulary, volume, effort, bugs, difficulty) & radon \\ \hline
Maintainability Index (MI) & radon \\ \hline
\end{tabular}
\end{table}
\vspace*{-\baselineskip}
\subsection{Maintainability Assessment Experiment}
\label{experiment}
%We show that these metrics can serve as early indicators of model modifiability and scalability by setting up the following experiment. 
%We make an assumption that the selected metrics can show a difference between two PyCATSHOO model designs at the early model development life cycle. 
The goal of our experiment is to  validate or refute the following hypothesis:
\begin{itemize}
\item{\textit{H1:\ }}The selected Maintainability metrics will show the difference between original and alternative model designs with respect to applied system level modifications;
\item{\textit{H2:\ }}The selected Maintainability metrics will show the difference between original and alternative model designs with respect to applied component level modifications;
\end{itemize}
We compare two designs of the Heated Room model reported in \cite{chraibi2013}: \\
%%%%%%%%%%%%%%%%%%%%%%%%%%%%%%%%%%%%%%%%%%%%%%
\textbf{Set 1: Original design.}
We take the original system model (Case 0) and create new versions of this model applying system level modifications (Case 1) and component level modifications (Case 2) defined above (see Table \ref{models}). These modifications illustrate an adaptive maintenance of a DPRA model in order to reflect real-size system requirements.
This set of three models is created following the original design ideas from \cite{chraibi2013} where system components are connected via PyCATSHOO message boxes, i.e., point-to-point.\\
%%%%%%%%%%%%%%%%%%%%%%%%%%%%%%%%%%%%%%%%%%%%%%
\textbf{Set 2: Alternative design.}
We create a set of semantically equivalent models of the Heated Room with alternative design (Case 0a - 2a). We promote the low coupling design principle by implementing well-known patterns of object-oriented design (i.e., Mediator, Observer) \cite{wolfgang1994design}). \\
%%%%%%%%%%%%%%%%%%%%%%%%%%%%%%%%%%%%%%%%%%%%%%
\textbf{Maintainability Assessment and Analysis of Results.}
We use the metrics from Table \ref{metrics} for both sets of models from Table \ref{models} and analyse the results. %In this work, we use only metrics supported by the tools. %OO-specific measures and Structural complexity measures were not considered.

In the following sections, we explain the Heated Room test case, provide the details on this experiment and discuss the maitainability assessment results.

%We show that the selected metrics can serve as early indicators of model modifiability and scalability. These indicators would allow experts to make better decisions early in the DPRA model development life cycle.  

\section{Case Study: Heated Room}

“Heated Room” is the test case reported in \cite{bouissou2013critical,bouissou2007comparison}.
This case is about a room which contains a heater device equipped with a thermostat. The latter switches the heater off when the room temperature reaches 22\deg C and switches it on when the room temperature
falls below 15\deg C. 
In this situation, a constant heat flow enters the room. The outside temperature is 13\deg C and at the initial time
t = 0 the room temperature is 17\deg C.
The flow of the heat leak through the walls is proportional to the difference between inside and outside temperatures. %The evolution of the room temperature can be described as follows:
%$$ dT/dt = leak * (T_{outside}-T)+ power_{heater}*heaterIsOn$$
The temperature is governed by a linear differential
equation as: $ dT/dt = \alpha T + \beta $ where $\alpha$ and $\beta$ depend on the mode of heater. The heater is assumed to have a constant failure rate
$\lambda = 0.01$ and repair rate $\mu =0.1$.
\subsection{Set 1: The Original PyCATSHOO Model of Heated Room}
\begin{figure}[H]   % the [b] specifies bottom of the page
\centering          % this centers everything inside the figure environment
\includegraphics[width=0.75\textwidth]{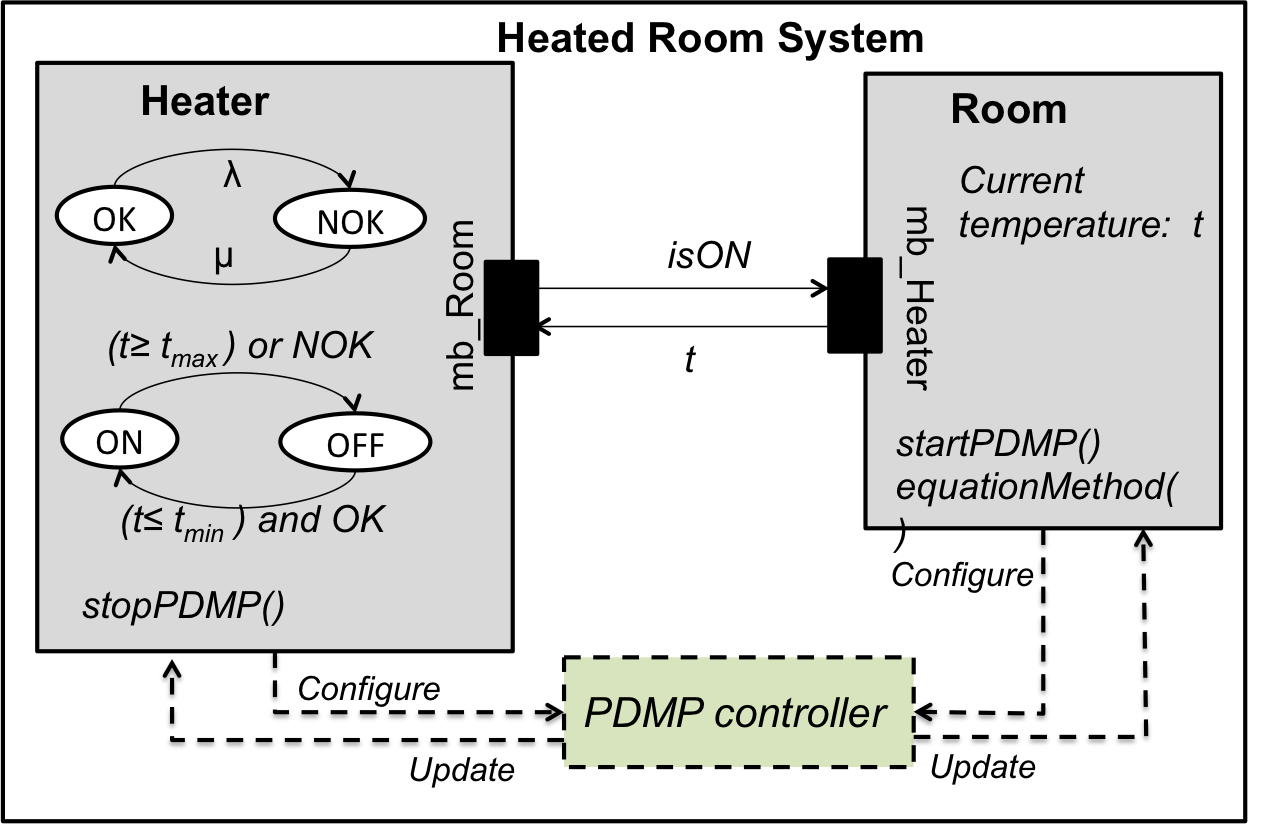}
\caption{Heated Room: The original model design using message boxes. Point-to-point connection between system components}
\label{fig:concept1}
\end{figure}

The original model of Heated Room is presented in \cite{chraibi2013}. It specifies the system components and their representing classes. The diagram in Fig. \ref{fig:concept1} shows the model from \cite{chraibi2013}. For the moment of this publication, the PyCATSHOO modeling tool does not have an explicit graphical modeling notation. We use the one that is adopted by EDF experts working with PyCATSHOO.
The Heater class defines two automata that describe functional and dysfunctional behavior of the heater. The OK/NOK transitions are stochastic: the time instants when the heater can fail and can be repaired are distributed according to the exponential distribution law with the parameters $\lambda$ and $\mu$ accordingly. The ON/OFF transitions follow the control logic: the heater switches ON if it is in the functional state (OK) and when the room temperature drops below its specified min. temperature $T_{min} = 15$ and switches OFF when it raises above the specified max. temperature $T_{max}=22$. The listing below describes the Heater class, its variables, automata and message boxes according to Fig. \ref{fig:concept1}
\begin{small}
\begin{lstlisting}[language=Python, caption={Definition of \textbf{Heater} class}, label={lst:heater}]
import Pycatshoo as Pyc
class Heater(Pyc.CComponent):
...
# variables
 self.po_maxTemperature = self.addVariable("maxTemperature", Pyc.VarType.double, 22)
 self.po_minTemperature = self.addVariable("minTemperature", Pyc.VarType.double, 15)
 self.po_power          = self.addVariable("power"         , Pyc.VarType.double, 1.0)
 self.po_lambda         = self.addVariable("lambda", Pyc.VarType.float, 0.01)
 self.po_mu             = self.addVariable("mu", Pyc.VarType.float, 0.1)
...
# heater states
 self.a1 = self.addAutomaton("Function") #Functioning
 self.stateOK  = self.addState("Function", "OK" , 1)
 self.stateNOK = self.addState("Function", "NOK", 0)
 self.setInitState("OK")
    
 self.a2 = self.addAutomaton("Power") #Control logic
 self.stateON  = self.addState("Power", "ON" , 1)
 self.stateOFF = self.addState("Power", "OFF", 0)
 self.setInitState("ON")
...
#heater transitions
 trans = self.stateOK.addTransition("OK_to_NOK")
 trans.setDistLaw(Pyc.IDistLaw.newLaw(self, Pyc.TLawType.expo, self.po_lambda))
...
 trans = self.stateNOK.addTransition("NOK_to_OK")
 trans.setDistLaw(Pyc.IDistLaw.newLaw(self, Pyc.TLawType.expo, self.po_mu))
...
 trans = self.stateOFF.addTransition("OFF_to_ON")
 trans.setDistLaw(Pyc.IDistLaw.newLaw(self, Pyc.TLawType.inst, 1))
 trans.setCondition("ONCondition",
 lambda: self.stateOK.isActive() and (self.getTemperature() <= self.po_minTemperature.dValue()), False)
...    
trans = self.stateON.addTransition("ON_to_OFF")
trans.setDistLaw(Pyc.IDistLaw.newLaw(self, Pyc.TLawType.inst, 1))
trans.setCondition("OFFCondition",
lambda: self.getTemperature() >= self.po_maxTemperature.dValue() or self.stateNOK.isActive(), False)  
... 
# Message boxes
 self.addMessageBox("mb_Room")
 self.addMessageBoxExport("mb_Room", self.stateON, "heaterON")
 self.addMessageBoxExport("mb_Room",self.po_power,"heatingPower")
 self.addMessageBoxImport("mb_Room",self.pi_roomTemperature,"temperature")
...
def stopCondition(self):
...
\end{lstlisting}
\end{small}
The Room class represents an observed subject: its temperature continuously evolves. %The current temperature value is observed by the heater and used in its control logic. 
The Heater and the Room communicate via message boxes: the Room component sends its current temperature to the heater, whereas the Heater sends its current state (ON or OFF) and its power value to the Room. 
\begin{lstlisting}[language=Python, caption={Definition of \textbf{Room} class}, label={lst:room}]
import Pycatshoo as Pyc
class Room(Pyc.CComponent):
...
# variables
...
 self.po_current_temperature = self.addVariable("temperature",   Pyc.VarType.double, 17.0)   
# Pycatshoo Inports
 self.pi_heaterIsOn = self.addReference("heaterON")
 self.pi_hPower = self.addReference("heatingPower")
# Message boxes
 self.addMessageBox("mb_Heater")
 self.addMessageBoxExport("mb_Heater", self.po_current_temperature, "temperature")
 self.addMessageBoxImport("mb_Heater", self.pi_heaterIsOn         , "heaterON")
\end{lstlisting}
In the original model design, the Room class contains the specification of a PDMP controller that implements the physics of the process - the evolution of the room temperature over time $T(t)$.
\begin{lstlisting}[language=Python, caption={Definition of \textbf{Room} class}, label={lst:room1}]
# Creating PDMP Controller for the System 
  self.PDMP_temp = "pdmpTemperature"
  self.system().addPDMPManager(self.PDMP_temp)
  self.addPDMPEquationMethod(self.PDMP_temp, "pdmpMethod", None, 0)
  self.addPDMPODEVariable(self.PDMP_temp, self.po_current_temperature)
  self.addStartMethod("start", self.start)
...

# only a single heater is considered
def pdmpMethod(self):
  if self.pi_heaterIsOn.bValue(0):
    power = self.pi_hPower.dValue(0)
  else:
    power = 0
  self.po_current_temperature.setDvdtODE(power - self.po_leakage.dValue() * (self.po_current_temperature.dValue() - self.po_outside_temperature.dValue()))
\end{lstlisting}
The PDMP controller is a part of the system. Other system components define the functioning of the PDMP controller in a distributed manner:  \textit{equationMethod()} for the room temperature  is defined  by the Room class, stop conditions \textit{stopPDMP()} (e.g., when the boundary temperature value is reached) are defined by the Heater class.   

\begin{figure}   % the [b] specifies bottom of the page
\centering          % this centers everything inside the figure environment
\includegraphics[width=0.99\textwidth]{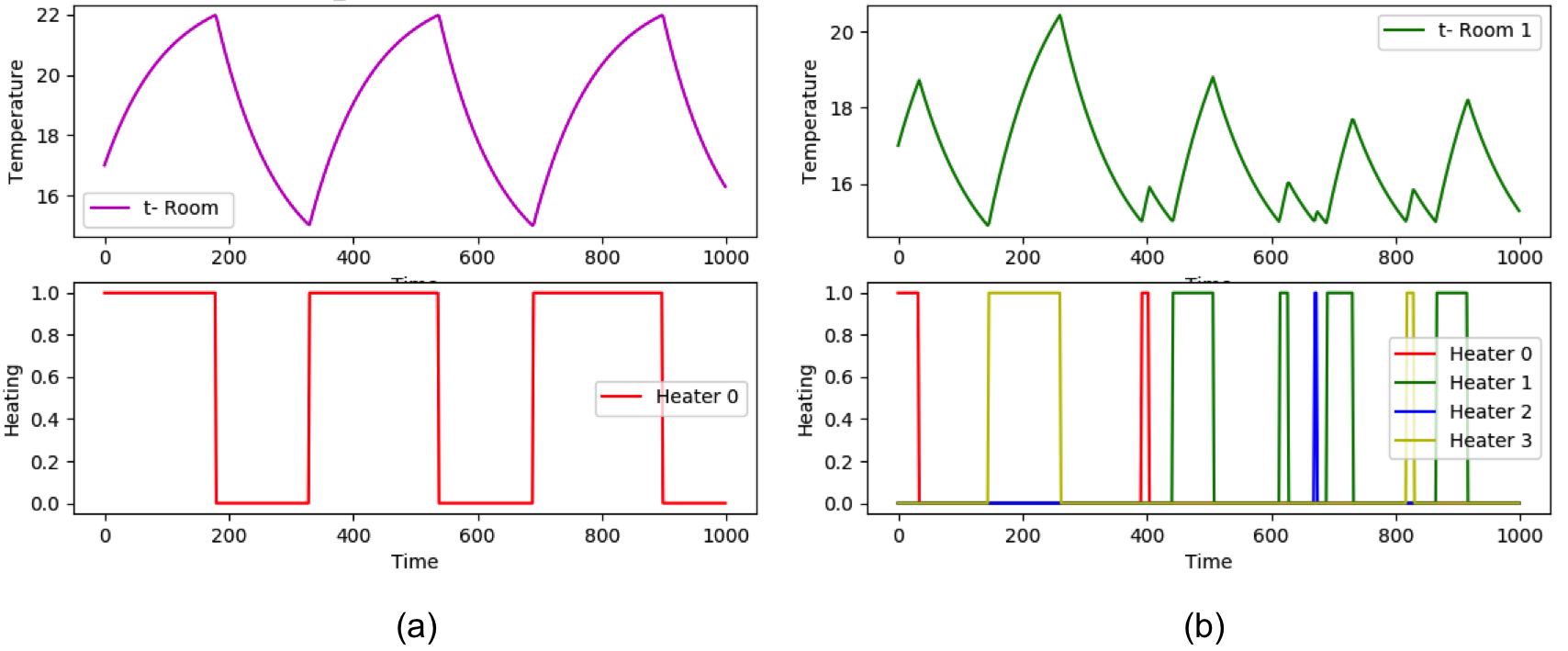}
\caption{Simulation results for PyCATSHOO model of the Heated Room system: a) Case 0/0a: 1 room, 1 heater: $\lambda = 0.001, \mu=0.01$; b) Case 2/2a: 1 room, 4 heaters with standby redundancy: H0,.. H3; H0: $\lambda = 0.02, \mu=0.01$; Backups H1..H3: $\lambda = 0.01, \mu=0.01$}
\label{fig:sym}
\end{figure}

In the initial model (Case 0), the Heated Room System class specifies the system with one heater and one room objects connected via corresponding message boxes (mb\_Room and mb\_Heater). 
%%%% PROBABLY THE LISTING??
\begin{lstlisting}[language=Python, caption={Original design of the \textbf{HeatedRoomSystem} class: Case 0}, label={lst:hr}]
class HeatedRoomSystem(Pyc.CSystem):
# Instantiation of heater and room
  self.heater = Heater("Heater")
  self.room   = Room("Room")
# connecting heater and room via message boxes:
  self.connect("Heater", "mb_Room", "Room", "mb_Heater")
\end{lstlisting}
%In the initial case (Case 0), the Heated Room system consists of 1 heater connected to 1 room. 
Fig.\ref{fig:sym}a illustrates the execution trace of the model with the PyCATSHOO simulator: the graph of temperature evolution over time  and the heating regime of the heater (ON/OFF).
We do not provide the Python code listing due to the space limitations. The detailed specification of the model is presented in \cite{chraibi2013}. 
\vspace*{-\baselineskip}
\subsection{Adaptive Maintenance of the Heated Room Model: Case 1, 2}
\label{HR_modif}
%Table \ref{cases} summarises the modifications applied to the initial PyCATSHOO model of the Heated Room system.\\
\textbf{System level modifications:}
Case 1 illustrates the system level modifications.  We modify the structure and topology of the Heated Room system by instantiating four heaters (H0..H3) and connecting them explicitly to the  room via message boxes. %We extend the system by adding three heaters. 

The Heater class does not change: the heater components independently heat the room following their initial control logic. Several heaters can be ON or OFF at the same time, based on the room temperature and their functioning state (OK or NOK).

We modify the Room class and generalize the PDMP equation method in order to establish the connection between the room and multiple heaters. 
\begin{lstlisting}[language=Python, caption={Modified \textbf{Room} class: Case 1}, label={lst:hr1}]
class Room(Pyc.CComponent):
...
# we consider an arbitrary number of heaters
def pdmpMethod(self):
    power = 0
    if (self.pi_heaterIsOn.nbCnx() > 0):
        # if multiple heaters - the power increases proportionally!
        power = 0
        for i in range(0, self.pi_heaterIsOn.nbCnx()):
            power = power + self.pi_hPower.dValue(i) * self.pi_heaterIsOn.bValue(i)
    self.po_current_temperature.setDvdtODE(power - self.po_leakage.dValue() * (
    self.po_current_temperature.dValue() - self.po_outside_temperature.dValue()))
\end{lstlisting}
The Heated Room System class specifies the system with four heaters and one room connected via corresponding message boxes (mb\_Room and mb\_Heater). 
%%%% PROBABLY THE LISTING??
\begin{lstlisting}[language=Python, caption={Original design of the \textbf{HeatedRoomSystem} class: Case 1}, label={lst:hr}]
class HeatedRoomSystem(Pyc.CSystem):
def __init__(self):
    Pyc.CSystem.__init__(self,"HeatedRoom")

# Instantiation of heaters and room ---
    self.room = Room("Room")
    self.heater0 = Heater("Heater0")
    self.heater1 = Heater("Heater1")
    self.heater2 = Heater("Heater2")
    self.heater3 = Heater("Heater3")
# connecting heater and room via message boxes:
    self.connect("Heater0", "mb_Room", "Room", "mb_Heater")
    self.connect("Heater1", "mb_Room", "Room", "mb_Heater")
    self.connect("Heater2", "mb_Room", "Room", "mb_Heater")
    self.connect("Heater3", "mb_Room", "Room", "mb_Heater")
\end{lstlisting}
\textbf{Component level modifications:}
Case 2 illustrates the component level modifications. We modify the control logic of the Heater component in order to support the standby redundancy mechanism: one heater (H0) will be declared as the main heater (the highest priority), whereas the other heaters will serve as its backups. Backup heaters are switching on only when all the other heaters with higher priority fail (NOK) and the temperature drops below $T_{min}$. If a heater with higher priority is repaired (OK), the backup heater with lower priority switches off. 
The following modifications are integrated into the Heater class:\\
- We add a variable for specifying heater's priority that will be communicated to other heaters and a corresponding message box:
\begin{lstlisting}[language=Python, caption={Standby redundancy implementation; Case 2}, label={lst:hr}]
...
# Initial Priority
 self.InitialPriority = MyPriority
# Priority of the heater
 self.po_priority = self.addVariable("priority", Pyc.VarType.int, MyPriority)
 self.addMessageBox("mb_OtherH_O")
 self.addMessageBoxExport("mb_OtherH_O", self.po_priority, "heaterPr")
\end{lstlisting}
- We add a variable and a corresponding message box to communicate with the other heaters (i.e., to receive their priority and status)
\begin{lstlisting}[language=Python, caption={}, label={lst:hr}]
...
self.pi_priorityOther = self.addReference("OtherHeatersPriority")
self.addMessageBox("mb_OtherH_I")
self.addMessageBoxImport("mb_OtherH_I", self.pi_priorityOther,"heaterPr")
\end{lstlisting}
- We modify the ON/OFF and OFF/ON transitions' conditions taking into account the priority of a current heaters and the statuses/priorities of the other heaters. For example, a heater H with priority x switches on only if all the heaters with priority $ y > x $ are off and the room temperature is below the threshold; conversely, the heater H with priority x switches off once a heater with priority $ y > x $ is switching on or once the room temperature is above the threshold.
\begin{lstlisting}[language=Python, caption={\textbf{Heater} class ON/OFF transition: Case 2}, label={lst:hr}]
  trans = self.stateON.addTransition("ON_to_OFF")
  trans.setDistLaw(Pyc.IDistLaw.newLaw(self, Pyc.TLawType.inst, 1))
  trans.setCondition("OFFCondition",
# NOK
    lambda: self.stateNOK.isActive() or
# T> max   
    (self.getTemperature() >= self.po_maxTemperature.dValue()) or 
# Another Heater with higher priority is OK   
    (self.getMyPriority() < self.getOtherPriority())  
   , False)
 ...
\end{lstlisting}
- We modify the stop condition for the PDMP controller taking into account the priority of a current heaters and the statuses/priorities of the other heaters\footnote{We omit the details of the \textit{getMyPriority()},\textit{ getOtherPriority()} methods as the standby redundancy can be implemented in various ways and discussion of a particular algorithm is out of our scope.}.

We  modify the Heated Room System class where new heaters are explicitly connected to the room and to each other via message boxes. 
\begin{lstlisting}[language=Python, caption={The \textbf{HeatedRoomSystem} class: Case 2. Communication between four heaters and a room is ensured by message boxes that connect each pair of individual components.}, label={lst:hr}]
class HeatedRoomSystem(Pyc.CSystem):
...
  self.room = Room("Room") # Instantiation of room
# Instantiation of heater
#Backups have different priority
  self.heater0 = Heater("Heater0", True, 10)
  self.heater1 = Heater("Heater1", True, 6)
  self.heater2 = Heater("Heater2", True, 4)
  self.heater3 = Heater("Heater3", True, 2)
# connecting heaters and room via message boxes:
  self.connect("Heater0", "mb_Room", "Room", "mb_Heater")
  self.connect("Heater1", "mb_Room", "Room", "mb_Heater")
  self.connect("Heater2", "mb_Room", "Room", "mb_Heater")
  self.connect("Heater3", "mb_Room", "Room", "mb_Heater")
# connecting heaters each others
# 0 <-> 1
  self.connect("Heater0", "mb_OtherH_O", 
  "Heater1", "mb_OtherH_I")
  self.connect("Heater1", "mb_OtherH_O", 
  "Heater0", "mb_OtherH_I")
# 0 <-> 2
  self.connect("Heater0", "mb_OtherH_O", 
  "Heater2", "mb_OtherH_I")
  self.connect("Heater2", "mb_OtherH_O", 
  "Heater0", "mb_OtherH_I")
# 0 <-> 3
  self.connect("Heater0", "mb_OtherH_O", 
  "Heater3", "mb_OtherH_I")
  self.connect("Heater3", "mb_OtherH_O", 
  "Heater0", "mb_OtherH_I")
#   1 <-> 2
  self.connect("Heater1", "mb_OtherH_O", 
  "Heater2", "mb_OtherH_I")
  self.connect("Heater2", "mb_OtherH_O", 
  "Heater1", "mb_OtherH_I")
#   1 <-> 3
  self.connect("Heater1", "mb_OtherH_O", 
  "Heater3", "mb_OtherH_I")
  self.connect("Heater3", "mb_OtherH_O", 
  "Heater1", "mb_OtherH_I")
#     2 <-> 3
  self.connect("Heater2", "mb_OtherH_O", 
  "Heater3", "mb_OtherH_I")
  self.connect("Heater3", "mb_OtherH_O", 
  "Heater2", "mb_OtherH_I")
\end{lstlisting}
The graph of temperature evolution over time  and the heating regime of the four heaters H0..H3 for Case 2 are shown in Fig.\ref{fig:sym}b.

\subsection{Set 2: The Alternative Design of Heated Room }
\begin{figure}   % the [b] specifies bottom of the page
\centering          % this centers everything inside the figure environment
\includegraphics[width=0.75\textwidth]{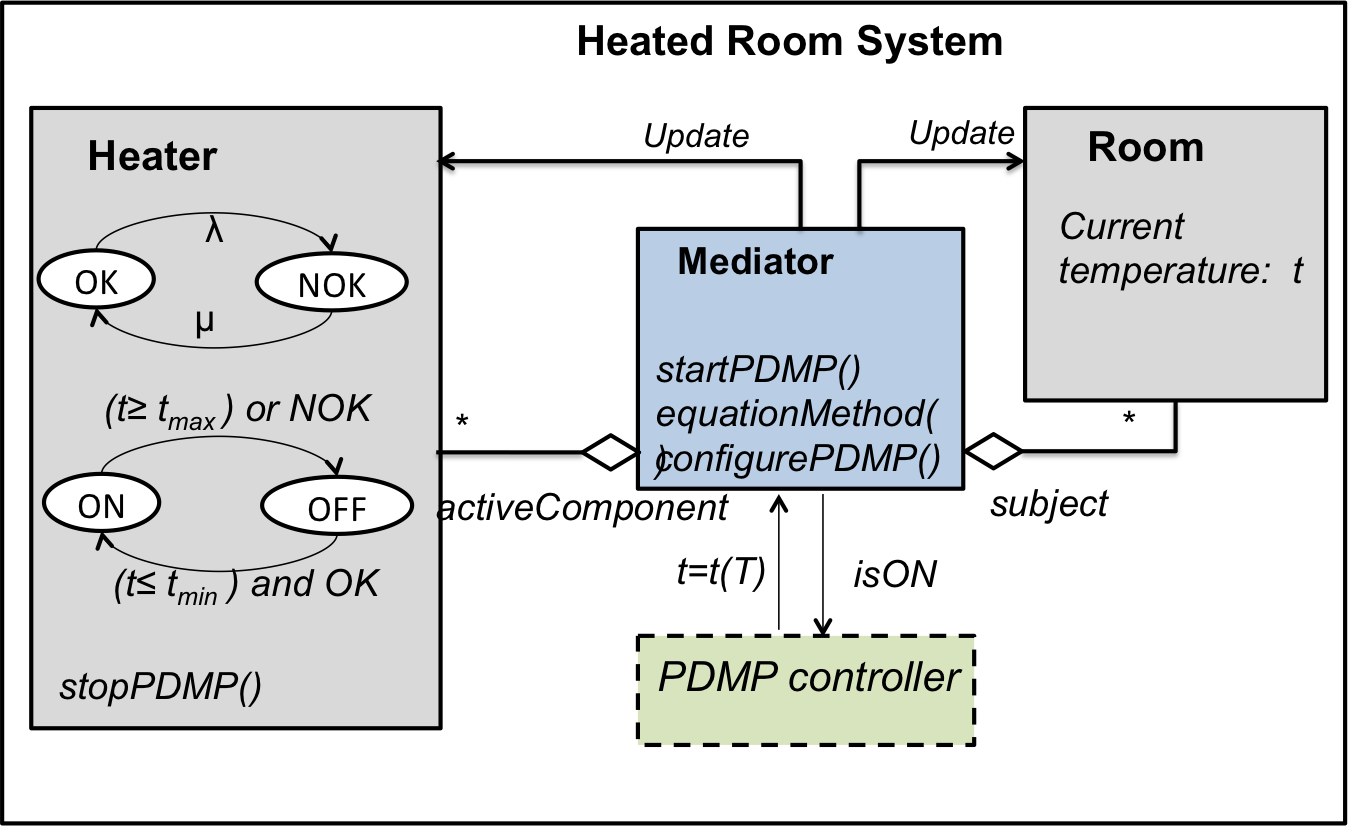}
\caption{Heated Room: The alternative model design using Design Patterns. Communication between components encapsulated by a Mediator}
\label{fig:concept2}
\end{figure}

We integrate the low-coupling design principles early in modeling phase in order to anticipate component level and system level modifications. We use well-known design patterns from OO software development \cite{wolfgang1994design}. The diagram in Fig. \ref{fig:concept2} shows the alternative design of the Heated room model.
The \textit{Mediator} design pattern encapsulates the interactions between system component and reduces direct dependencies between them. Integrating mediator early in the model improves system scalability.

In our design, the Mediator class maintains the lists of active components (heaters in our case) and subjects or passive components (i.e., rooms) as shown in Fig. \ref{fig:concept2}. It mediates the communication between the PDMP controller, the heater(s) and the room(s) replacing the point to point connections via message boxes. Note, that arbitrary number of rooms and heaters per room can be configured with this design. 
\begin{lstlisting}[language=Python, caption={Definition of abstract \textbf{Mediator} class}, label={lst:med}]
# A generic mediator class
class AbstractMediator(Pyc.CComponent):
def __init__(self, mediator_name, pdmp_ref, subjects_vars, subjects_objects ):
# list of active components in the system
 self.activeComponents = [] 
# list of subjects: {subject : ode_Var}
 self.subjects = {}  
# subjects and their active components are grouped 
 self.componentGroups = {}  # { subject : {component:type}}
...
# creating and configuring a PDMD manager in the system
 self.system().addPDMPManager(pdmp_ref)

def configurePDMP(self):
# START CONDITION
 self.addStartMethod(self.startControllerName, self.start)

# add the observable variable(s)  
# from all the subjects
 for v in self.subjects.values():
    self.addPDMPODEVariable(self.pdmp_ref, v)
# add the evaluation method 
# (as a function of active components)
 self.addPDMPEquationMethod(self.pdmp_ref, self.pdmpMethodName, self.pdmpEqMethod, 0)

# add the conditions to stop the PDMP 
# (an aggregate from all the active component)
 for c in self.activeComponents:
  c.addPDMPCondition(self.pdmp_ref, c.stopControllerName, c.stopCondition)
...
\end{lstlisting}
The mediator contains the \textit{configurePDMP()} method that allows to "assemble" the PDMP behavior from the parts defined by the active and passive components (i.e., \textit{startPDMP(), stopPDMP(), equationMethod()}). 
\begin{lstlisting}[language=Python, caption={Definition of the concrete \textbf{Mediator} class}, label={lst:med1}]
# A concrete Mediator for the Heated Room system:
class MultiRoomMediator(AbstractMediator):
...
# equation method that implements 
# a linear differential equation parametrised
# by multiple heaters
def pdmpEqMethod(self):
 for s, components in self.componentGroups.items():  
# s = subject in the list of subjects;
    heating_power = 0
    for c, type in components.items():  
# c = an active component assigned to s
    if (type == "heater"):
      if (c.stateON.isActive()):
        heating_power = heating_power + c.po_power.dValue()
    s_odeVar = self.subjects[s]  
# variable to evolve in the room s
    s_odeVar.setDvdtODE(heating_power - s.po_leakage.dValue()* 
    (s_odeVar.dValue() - s.po_outside_temperature.dValue()))
# once the room temperature is updated - we are updating 
# all the active components in this room 
# (ALTERNATIVE TO MESSAGE BOX)
    self.updateActiveComponents(components.keys(), s_odeVar.dValue()) 
    
# Update method:
def updateActiveComponents(self, components, newValue):
 for c in components:
  c.updateObservableVar(newValue)
\end{lstlisting}
%(e.g., stop conditions defined by different heaters, PDMP equation method(s) defined by the room(s) etc.). 
Note that different types of active components can be used once they do not change the type of the differential equation: for example, coolers can be used along with heaters as they only change the heating coefficient.  

The Heater class is similar to the original design; the message boxes are  removed. The Mediator object updates the heaters with the new value of the room temperature (lines 23, 26-28 in listing \ref{lst:med1}).

The Room class contains a current temperature variable that is updated by the Mediator component. Compared to the original design, the PDMP controller specification is moved from the Room class to the Mediator class. This allows to decouple the room and the heater.%: once the number of heaters or their behavior changes, this does not require modification in the Room class. 

In the initial model (Case 0a), the Heated Room System class specifies the system with one heater and one room attached to a mediator object.  \textit{configurePDMP()} method of the mediator terminates the system configuration. 
\begin{lstlisting}[language=Python, caption={ Alternative design of the \textbf{HeatedRoomSystem} class:  Case 0a}, label={lst:hr}]
class HeatedRoomSystem(Pyc.CSystem):
    # Instantiation of heater, room, mediator:
    self.heater = Heater("Heater") 
    self.room = Room("Room")
    self.heatedRoomMediator = Mediator("Mediator", 
        "pdmpTemperature", {self.room: self.room.temp},
        {self.heater: "heater"})
    #configuring PDMP controller:
    self.heatedRoomMediator.configurePDMP()
\end{lstlisting}
Simulation results of the Case 0a  are shown in Fig.\ref{fig:sym}a.
\vspace*{-\baselineskip}
\subsection{Adaptive Maintenance of the Alternative Model: Case 1a, 2a}
\textbf{System level modifications:}
Case 1a is an alternative design of the Heated Room system with 4 independent heaters. The Heater, the Room and the Mediator classes are not changed.

In the Heated Room System class new heaters are instantiated and added to the mediator.
\begin{lstlisting}[language=Python, caption={Alternative design of the \textbf{HeatedRoomSystem} class:  Case 1a}, label={lst:hr}]
class HeatedRoomSystem(Pyc.CSystem):
...
# Instantiation of heater and room
  self.room = Room("Room")
  self.heater0 = Heater("Heater0")
  self.heater1 = Heater("Heater1")
  self.heater2 = Heater("Heater2")
  self.heater3 = Heater("Heater3")
#Mediator
  self.heatedRoomMediator.addGroup(
   self.room, self.room.po_current_temperature,
   {self.heater0: "heater", self.heater1: "heater",
   self.heater2: "heater", self.heater3: "heater"})
#configurign PDMP controller:
  self.heatedRoomMediator.configurePDMP()
\end{lstlisting}
\textbf{Component level modifications:}
Case 2a implements the standby redundancy algorithm for heaters. We modify the Heater class following the logic described in Case 2 in section \ref{HR_modif}. Similar to the Case 2, we integrate the following modifications into the Heater class:\\
- We add a variable to specify the priority of the heater; \\
- We modify the ON/OFF and OFF/ON transitions' conditions taking into account the priority of a current heaters and the statuses/priorities of the other heaters;
\begin{lstlisting}[language=Python, caption={\textbf{Heater} class ON/OFF transition:  Case 2a}, label={lst:hr}]
class Heater(AbstractComponent):
...
self.takeON = self.addVariable("takeON", Pyc.VarType.bool, isMain)
...
#heater transitions
...
  trans = self.stateON.addTransition("ON_to_OFF")
  trans.setDistLaw(Pyc.IDistLaw.newLaw(self, Pyc.TLawType.inst, 1))
  trans.setCondition("OFFCondition",
   lambda: self.stateNOK.isActive() # failed
or
  (self.observableVar.dValue() >= self.po_maxTemperature.dValue()) # T > Tmax
or
  (not self.takeON.bValue()) # low priority
, False) 
\end{lstlisting}
- We modify the stop condition for the PDMP controller.\\
Compared to the original design, we do not connect heaters via message
boxes, but implement the Observer design pattern. 

The \textit{Observer} design pattern allows for implementing specific behavior between a group of system components. It defines a one-to-many dependency between objects and when one system component changes state, all its dependents are notified and updated automatically. 

The listing below illustrates a fragment of Observer implementation: we add a sensitive method called \textit{notifyFailure()} to the OK/NOK transition that executes each time the component fails (i.e., goes to the NOK state). This method notifies all the observers (i.e., heaters with lower priority or backup heaters) of the current heater that it fails and that they can now take on the heating. The \textit{takeON} variable of the corresponding observers is updated.  Along those lines, the observers are notified about the reparation or the switching on of their observable heater. We omit the logic of the algorithm.
\begin{lstlisting}[language=Python, caption={Implementing Observer pattern}, label={lst:hr}]
class Heater(AbstractComponent):
...
trans.addTarget(self.stateOFF, Pyc.TransType.trans)
...        
    trans = self.stateOK.addTransition("OK_to_NOK")
    trans.setDistLaw(Pyc.IDistLaw.newLaw(self, Pyc.TLawType.expo, self.po_lambda))
# when fail --> notifies possible dependents (backups) 
    trans.addSensitiveMethod("Failure", self.notifyFailure)
    trans.setCondition(True)
    trans.addTarget(self.stateNOK, Pyc.TransType.fault)    
...
def notifyFailure(self):  
# updating "takeON" status for all the backups in the chain 
  for next in self.backups:
    next.takeON.setBValue(True)  
    next.notifyFailure() 
\end{lstlisting}
In the Heated Room System class new heaters are instantiated, grouped according the main/backup topology and added to the mediator:
\begin{lstlisting}[language=Python, caption={Alternative design of the \textbf{HeatedRoomSystem} class:  Case 2a}, label={lst:hr}]
class HeatedRoomSystem(Pyc.CSystem):
...
# Instantiation of heater and room
  self.room = Room("Room")
  self.heater0 = Heater("Heater0", True) #main heater
  self.heater1 = Heater("Heater1", False)
  self.heater2 = Heater("Heater2", False)
  self.heater3 = Heater("Heater3", False)
# configuring backups  
  self.heater0.addBackup(self.heater1)
  self.heater1.addBackup(self.heater2)
  self.heater2.addBackup(self.heater3)
#Mediator (no change with Case 1a)
  self.heatedRoomMediator.addGroup(
   self.room, self.room.po_current_temperature,
   {self.heater0: "heater", self.heater1: "heater",
   self.heater2: "heater", self.heater3: "heater"})
#configurign PDMP controller:
  self.heatedRoomMediator.configurePDMP()   
\end{lstlisting}
The graphs of temperature evolution over time  and the heating regime of the four heaters H0..H3 for Case 2a are shown in Fig.\ref{fig:sym}b.

%In the initial case (Case 0a), the Heated Room system consists of 1 heater connected to 1 room. 

\section{Maintainability Assessment: Results}
We test the hypothesis from section \ref{experiment} applying the metrics to the model sets representing the original and the alternative designs. In this paper, we report on application the following metrics:
Line of Code (LOC); RLOC (relative modifications), Cyclomatic Complexity (CC); Halstead's volume, difficulty, vocabulary, effort, estimated bugs (V, D, $\eta$, E, B); compound measure Maintainability index (MI). These metrics are computed statically from the code.
\subsection{Size Metrics}
\subsubsection{LOC} (Lines of Code) is a software metric used to measure the size of a computer program. It is recognised as a valid indicator of complexity and maintainability. We use LOC as a measure of the size of the PyCATSHOO model.  The results of LOC measurement for two sets of PyCATSHOO models (Case 0-2, Case 0a-2a) are shown in Fig.\ref{fig:loc}. Stacked columns illustrate LOC per case; various colors correspond to various components. The following can be observed: \\
-  the Mediator class in Set 2 (alternative design) doubles the size of the simplified model;\\
- For the Set 1, the size of all the classes is growing in response to system and component level modifications; \\
- For the Set 2, only the Heater and the Heated Room System classes are growing. \\
Fig. \ref{fig:loc_t} shows the evolution of total LOC (model size) per case. Whereas for the initial model the alternative design results in a bigger model,  Case 2 and Case 2a models show almost equivalent size.
\subsubsection{RLOC.} 
\begin{figure}[H]
\begin{subfigure}{.5\textwidth}
  \centering
  \includegraphics[width=.9\linewidth]{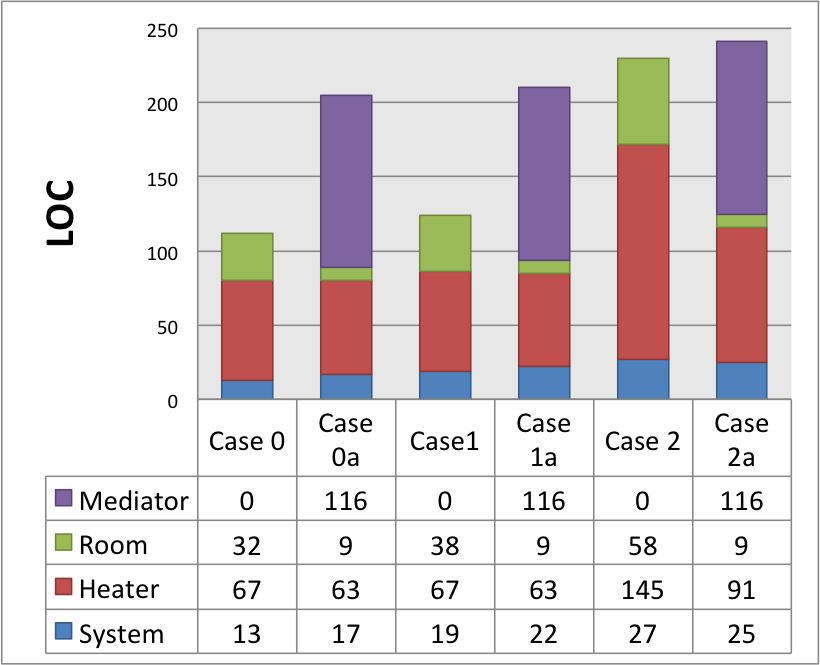}
  \caption{LOC analysis}
  \label{fig:loc}
\end{subfigure}%
\begin{subfigure}{.5\textwidth}
  \centering
  \includegraphics[width=0.9\linewidth]{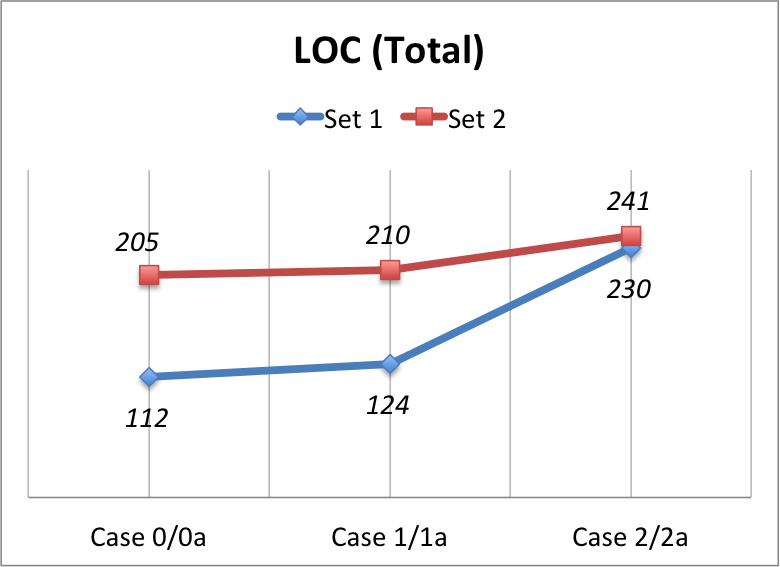}
  \caption{Evolution of total LOC per case}
  \label{fig:loc_t}
\end{subfigure}
\caption{}
\label{fig:diagrams}
\end{figure}

We propose to measure relative modifications (RLOC) in response to system and component modifications. We use cloc tool to measure the difference between pairs of models (Case 0 and Case 1; Case 1 and Case 2) in both model sets. We calculate total modification as a sum of modified, added and removed lines in the "adapted" model. 
We define RLOC for a case as follows: 
$$ RLOC_{case} = { LOC_{modif}+LOC_{add}+LOC_{rem} \over LOC_{case}}$$ 
This metric allows for more precise measurement of modifications as it takes into consideration not only the total change in model size. Removing or modifying the code (as well as model elements in concept model) are also considered.  Table \ref{rloc} summarises the RLOC measures. RLOC for the Set 1 shows that the model was modified on 82.25\% to implement scalability (from 1 to 4 heaters) and further on 108.69\% to implement standby redundancy of the heaters. \\
While having bigger initial size (Fig. \ref{fig:loc}), the model in the Set 2 was modified on 18.57\% to scale for 4 heaters and on 83.4\% to implement standby redundancy.
%\vspace*{-\baselineskip}
\begin{table}[]
\centering
\caption{RLOC comparison: Original vs. Alternative design}
\label{rloc}
\begin{tabular}{|@{ }l@{ }|@{ }l@{ }|@{ }l@{ }|@{ }l@{ }|}
\hline
\rowcolor[HTML]{EFEFEF} 
\multicolumn{1}{| c |}{\cellcolor[HTML]{EFEFEF}\textbf{Set 1}} & \textbf{RLOC} & \multicolumn{1}{| c |}{\cellcolor[HTML]{EFEFEF}\textbf{Set 2}} & \textbf{RLOC } \\ \hline
$Case 0 \longrightarrow Case 1$ &  \cellcolor[HTML]{EFEFEF}82,25\% & 
$Case 0a \longrightarrow Case 1a $ &  \cellcolor[HTML]{EFEFEF}18,57\% \\ \hline
$Case 1 \longrightarrow Case 2$  &  \cellcolor[HTML]{EFEFEF}108,69\% & 
$Case 1a \longrightarrow Case 2a$  & \cellcolor[HTML]{EFEFEF}83,4\% \\ \hline
\end{tabular}
\end{table}
%%%%%%%%%%%%%%%%%%%%%%%%%%%%%%%%%%%%%%%%%%%%%%%%%%%%%%
%\vspace*{-\baselineskip}
%\vspace*{-\baselineskip}
%%%%%%%%%%%%%%%%%%%%%%%%%%%%%%%%%%%%%%%%%%%%%%%%%%%%%%%%%%%%%
\subsection{Cyclomatic Complexity}
\begin{figure}[H]   % the [b] specifies bottom of the page
\centering          % this centers everything inside the figure environment
\includegraphics[width=0.6\textwidth]{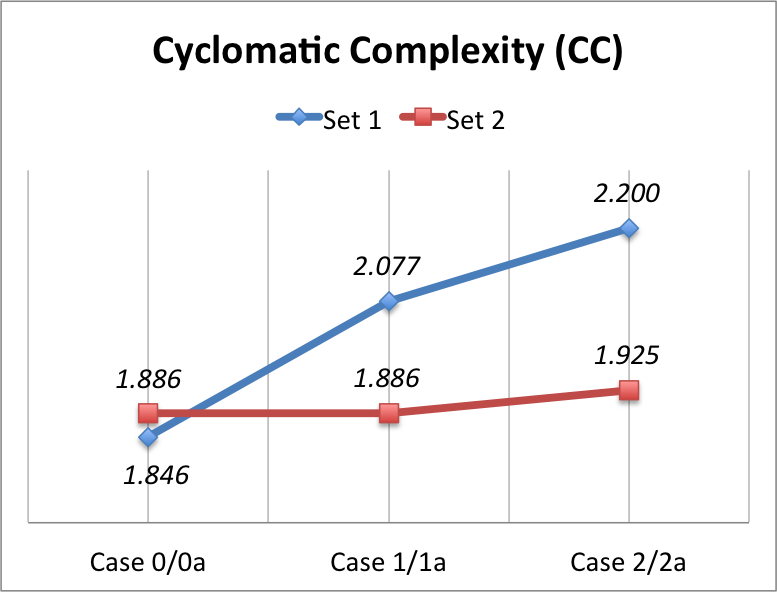}
\caption{Cyclomatic Complexity analysis}
\label{fig:cc}
\end{figure}

McCabe \cite{mccabe1976complexity} proposed Cyclomatic Complexity Measure to quantify complexity of a given software based on its flow-graph. A flow graph is based on decision-making constructs of a program. %The underlying assumption is: given two program of the same size, the one with more decision-making statements will be more complex as the control of program jumps frequently.
Fig.\ref{fig:cc} illustrates cyclomatic complexity measure for the model sets. Whereas the absolute CC values for all models indicate low complexity (below 10), we are interested in the CC evolution in response to adaptive model modifications. The graph in Fig.\ref{fig:cc} indicates the faster complexity growth for the Set 1. This corroborates with the previous measures.
%\vspace*{-\baselineskip}
%\vspace*{-\baselineskip}
\subsection{Halstead lexical measures}
\begin{figure}[]
\begin{subfigure}{.5\textwidth}
  \centering
  \includegraphics[width=0.9\linewidth]{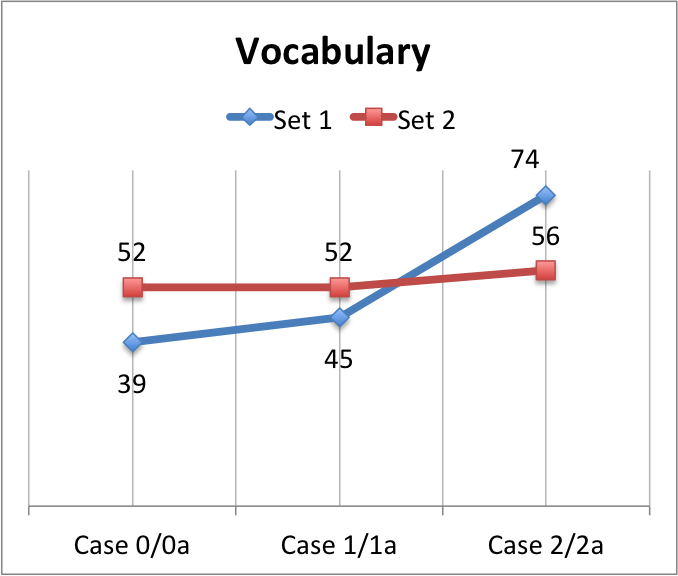}
  \caption{Vocabulary}
  \label{fig:H_voc}
\end{subfigure}%
\begin{subfigure}{.5\textwidth}
  \centering
  \includegraphics[width=.9\linewidth]{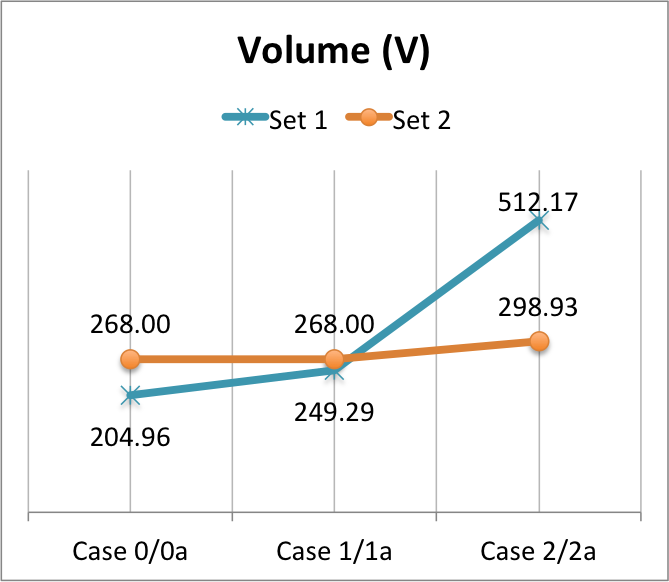}
  \caption{Volume}
  \label{fig:H_v}  
\end{subfigure}
\begin{subfigure}{.5\textwidth}
  \centering
  \includegraphics[width=.9\linewidth]{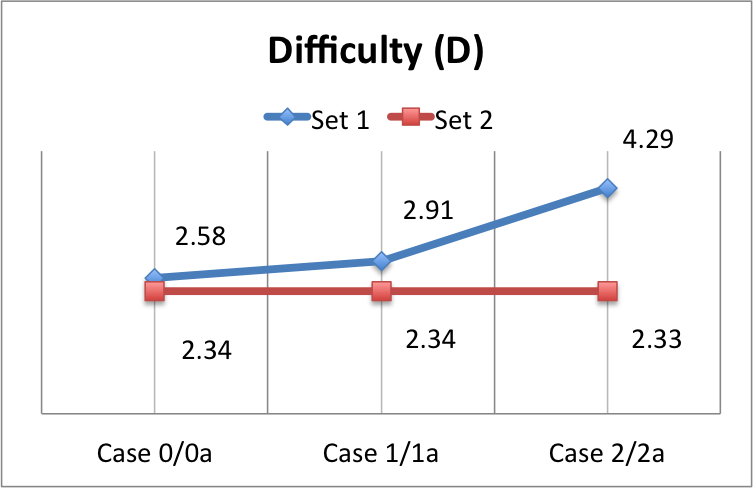}
  \caption{Difficulty}
  \label{fig:H_d}
\end{subfigure}%
\begin{subfigure}{.5\textwidth}
  \centering
  \includegraphics[width=0.9\linewidth]{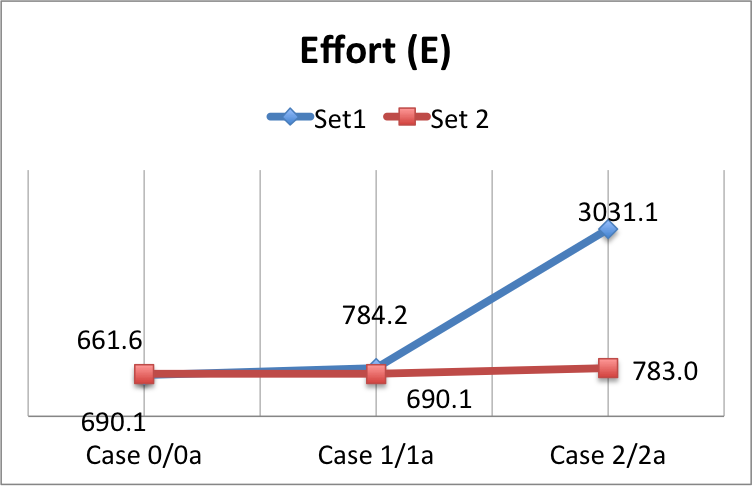}
  \caption{Effort}
  \label{fig:H_e}
\end{subfigure}
\begin{subfigure}{.5\textwidth}
  \centering
  \includegraphics[width=.9\linewidth]{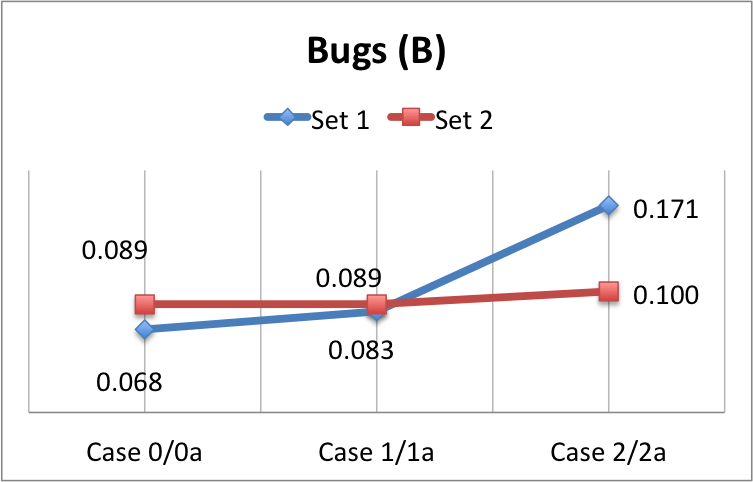}
  \caption{Estimated erors}
  \label{fig:H_b}
\end{subfigure}
\caption{}
\label{fig:H1}
\end{figure}

Halstead metrics consider a program as sequence of operators and their associated operands. 
For a given problem, let: \\
$\eta _{1}$  = the number of distinct operators; \\
$\eta _{2}$  = the number of distinct operands; \\
$N_{1}$ = the total number of operators; \\
$N_{2}$ = the total number of operands.\\
From these numbers, several measures can be calculated. In this work, we studied 
Program length $N$, vocabulary $\eta$, volume $V$, difficulty $D$, effort $E$ and estimated number of errors (or bugs) $B$:
$$ N = N_{1} + N_{2} $$
$$\eta =\eta _{1}+\eta _{2}$$ 
$$ V=N\times \log _{2}\eta$$
$$D={\eta _{1} \over 2}\times {N_{2} \over \eta _{2}} $$
$$E=D\times V $$
$$B={V \over 3000}$$
Fig.\ref{fig:H_b} illustrates some of Halstead metrics for two modeling sets and an evolution of these metrics in response to adaptive model modifications. 

Vocabulary measure is calculated as a sum of distinct operands and operators and can be related to the model complexity. Intuitively, a model with a bigger vocabulary will need more adaptive modifications in response to changed requirements. Thus, it is more difficult to maintain.

The volume measure illustrated in Fig.\ref{fig:H_v} shows the evolution of model volume between cases. Note that the results do not correspond to the evolution of model size measured with LOC in Fig. \ref{fig:loc_t}. It can be explained by the fact that the volume measure takes into account meaningful operands rather lines of code. 

The difficulty measure can be related to the difficulty of the PyCATSHOO model to write or to understand. The effort measure can be translated to an estimated development time and thus can indicate the cost of model for a PyCATSHOO model.
Delivered bugs is an estimate for the number of errors.
For the Case 0a, an estimate number of errors is higher due to implementation of the Mediator; however it seems to remain stable after adaptive modifications (Case 1a, Case 2a). Idem for the difficulty and effort measures.
%%%%%%%%%%%%%%%%%%%%%%%%%%%%%%%%%%%%%%%%%%%%%%%%%%%%%%
%\vspace*{-\baselineskip}
%\vspace*{-\baselineskip}
%%%%%%%%%%%%%%%%%%%%%%%%%%%%%%%%%%%%%%%%%%%%%%%%%%%%%%%%%%%%%
\subsection{Maintainability Index}
\begin{figure}   % the [b] specifies bottom of the page
\centering          % this centers everything inside the figure environment
\includegraphics[width=0.95\textwidth]{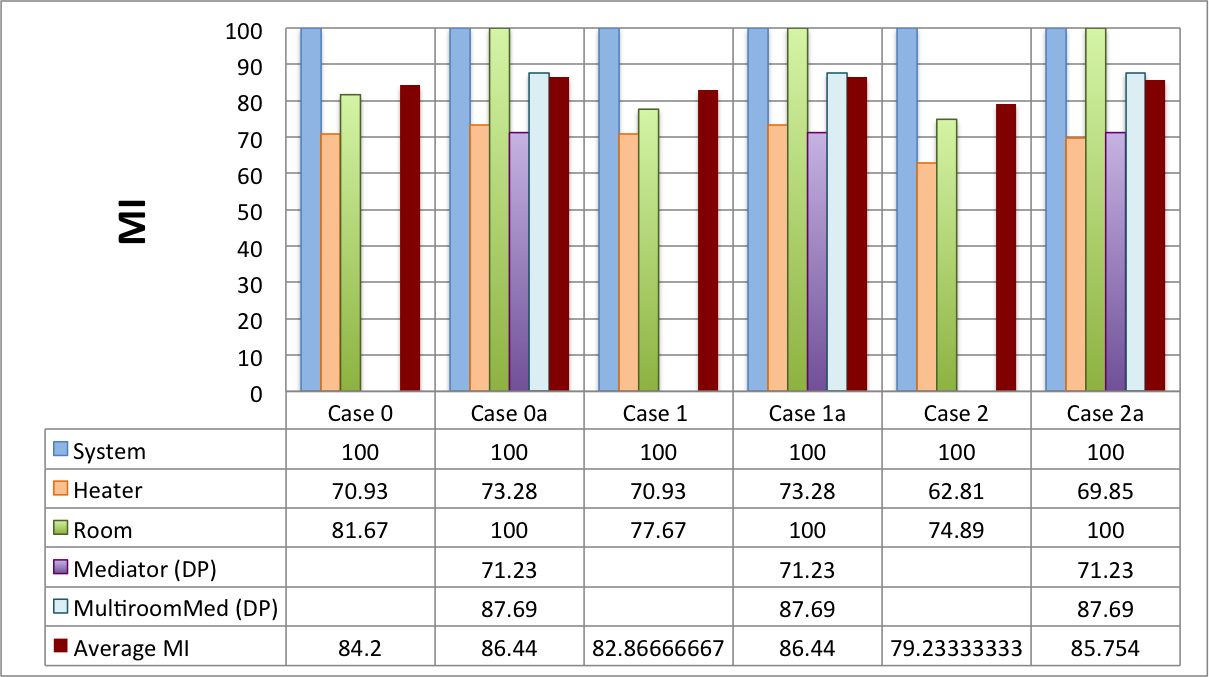}
\caption{Maintainability Index analysis}
\label{fig:mi}
\end{figure}

Maintainability Index is a software metric which measures how maintainable (easy to support and change) the source code is. The maintainability index is calculated as a factored formula consisting of Lines Of Code, Cyclomatic Complexity and Halstead volume:
$$ MI = 171 - 5.2 \times \ln(V) - 0.23 \times CC - 16.2 \times \ln(LOC) $$
Where V is a Halstead Volume; CC - Cyclomatic Complexity; LOC = count of source Lines Of Code. Fig.\ref{fig:mi} illustrates MI measure for the model sets: \\
-Room class of the Set 2 and Heated Room System class of both sets are considered as 100\% maintainable. \\
-MI of the Heater class decreases for both sets. This can be explained by its growing complexity and size. \\
-Average MI, though it remains very high for both sets, drops from 84.2 to 79.2 for the Set 1 and from 86.4 to 85.75 for the Set 2.
%\vspace*{-\baselineskip}
%\vspace*{-\baselineskip}
%\vspace*{-\baselineskip}
\section{Conclusion}

In this work, we proposed a pragmatic approach for assessing maintainability of DPRA models created with PyCATSHOO based on measures accepted in SE and OO design (LOC, CC, Halstead, MI). We claim that the selected metrics can be used as early indicators for estimating complexity, modifiability and, as a consequence, maintainability of the PyCATSHOO models. To validate our hypothesis, we compare two designs of the "Heated Room" system. 

We create two model sets, each consisting of three models: the initial model with 1 heater and 1 room and the two models, where we apply system level and component level modifications respectively.
Models in Set 1 follow the original design reported in \cite{chraibi2013} and based on point to point connection between system components. Set 2 promotes the low-coupling design principles. 
We use the selected metrics on both sets of models and analyse the results. %In this work, we use only metrics supported by the tools.
We also define our own metric - RLOC - that provides more accurate measure of model modification compared to LOC.

The results seem to validate our hypothesis: the selected maintainability metrics show the difference between original and alternative model designs at early model development phase and thus can help the experts to make design decisions.
 
Our results demonstrate the following: 

The original PyCATSHOO model of Heated room uses direct (point-to-point) connection between its components. Whereas this makes the model clear and easy to create for the case of two components, its complexity grows faster when the number of components increases or when new types of dependencies between components are introduced. 

The PyCATSHOO model based on design patterns requires more efforts at creation. In particular, we make some assumptions about how the system will evolve (its size, the types of dependencies) and propose some anticipated design solutions that ensure scalability along the considered dependencies "by design". this can be considered as an "upfront investment" into the maintainability.%As a result, the model based on design patterns demonstrates better maintainability and lower complexity compared to the original model. 
The evaluation indicates that the low coupling design principle applied early in the modeling improves scalability and maintainability of the real-size model for DPRA. So, the upfront investment pays off.
\subsection{Directions for the Future Work}
Despite the encouraging results, we consider them as preliminaries. We consider the following directions for our future work:
\begin{itemize}
    \item We plan to replicate our experiment and to assess the maintainability of the real PyCATSHOO projects. The proposed set of measures can be applied to available libraries of PyCATSHOO components (so called PyCATSHOO Knowledge Base or PKB). The results can be used to estimate the current complexity and maintainability of the models and to give an insight about the current design decisions and best practices adopted at EDF.
    \item We plan to explore the executable nature of the PyCATSHOO models and to integrate the metrics based on call graph, similar to \cite{bhattacharya2012graph}. The metrics based on call graph topology can be used to estimate the model evolution and performance.
    \item We plan to extend the list of metrics integrating the metrics on Structural complexity and other OO-specific metrics. The metrics discussed in \cite{rizvi2010maintainability,genero2005survey} are of a great interest and can complement the software measures studied in this work.  For the initial PyCATSHOO models or toy examples, the number of classes, dependencies and the size of hierarchies are small; as a result of adaptive maintenance and a transformation to a real-size model, the number and variety of components and dependencies between them grow bigger - here the specific OO-metrics can indicate the growth in size and complexity the conceptual model of a system, provide an insight on the element cohesion, coupling and help assessing the models understandability.
\end{itemize}

\bibliographystyle{splncs03}
\bibliography{rychkova_biblio}

\begin{thebibliography}{10}
\providecommand{\url}[1]{\texttt{#1}}
\providecommand{\urlprefix}{URL }

\bibitem{e1995mood}
Brito~e Abreu, F.: The mood metrics set. In: proc. ECOOP. vol.~95, p. 267
  (1995)

\bibitem{aldemir2013survey}
Aldemir, T.: A survey of dynamic methodologies for probabilistic safety
  assessment of nuclear power plants. Annals of Nuclear Energy  52,  113--124
  (2013)

\bibitem{bhattacharya2012graph}
Bhattacharya, P., Iliofotou, M., Neamtiu, I., Faloutsos, M.: Graph-based
  analysis and prediction for software evolution. In: Software Engineering
  (ICSE), 2012 34th International Conference on. pp. 419--429. IEEE (2012)

\bibitem{bouissou2013critical}
Bouissou, M., Chraibi, H., Chubarova, I.: Critical comparison of two user
  friendly tools to study piecewise deterministic markov processes (pdmp):
  season 2. ESREL (2013)

\bibitem{bouissou2007comparison}
Bouissou, M.: Comparison of two monte carlo schemes for simulating piecewise
  deterministic markov processes. Proceedings of Mathematical Methods in
  Reliability MMR  (2007)

\bibitem{cherfi2002conceptual}
Cherfi, S.S.S., Akoka, J., Comyn-Wattiau, I.: Conceptual modeling quality-from
  eer to uml schemas evaluation. In: International Conference on Conceptual
  Modeling. pp. 414--428. Springer (2002)

\bibitem{chidamber1994metrics}
Chidamber, S.R., Kemerer, C.F.: A metrics suite for object oriented design.
  IEEE Transactions on software engineering  20(6),  476--493 (1994)

\bibitem{chraibi2013}
Chraibi, H.: Dynamic reliability modeling and assessment with pycatshoo:
  Application to a test case. PSAM (2013)

\bibitem{chraibi2016}
Chraibi, H.: Pycatshoo:toward a new platform dedicated to dynamic reliability
  assessments of hybrid systems. PSAM (2016)

\bibitem{DPRA_Difficulty}
Coyne, K., Siu, N.: {Simulation-Based Analysis for Nuclear Power Plant Risk
  Assessment: Opportunities and Challenges}. In: Proceeding of the ANS Embedded
  Conference on Risk Management for Complex Socio-Technical Systems (2013)

\bibitem{genero2007building}
Genero, M., Manso, E., Visaggio, A., Canfora, G., Piattini, M.: Building
  measure-based prediction models for uml class diagram maintainability.
  Empirical Software Engineering  12(5),  517--549 (2007)

\bibitem{genero2002defining}
Genero, M., Miranda, D., Piattini, M.: Defining and validating metrics for uml
  statechart diagrams. Proceedings of QAOOSE  2002 (2002)

\bibitem{genero2000early}
Genero, M., Piattini, M., Calero, C.: Early measures for uml class diagrams.
  L’objet  6(4),  489--505 (2000)

\bibitem{genero2002empirical}
Genero, M., Piattini, M., Calero, C.: Empirical validation of class diagram
  metrics. In: Empirical Software Engineering, 2002. Proceedings. 2002
  International Symposium n. pp. 195--203. IEEE (2002)

\bibitem{genero2005survey}
Genero, M., Piattini, M., Calero, C.: A survey of metrics for uml class
  diagrams. Journal of object technology  4(9),  59--92 (2005)

\bibitem{gilb2008designing}
Gilb, T.: Designing maintainability in software engineering: a quantified
  approach. International Council on Systems Engineering (INCOSE)  (2008)

\bibitem{halstead1977elements}
Halstead, M.H.: Elements of software science, vol.~7. Elsevier New York (1977)

\bibitem{harrison1998evaluation}
Harrison, R., Counsell, S.J., Nithi, R.V.: An evaluation of the mood set of
  object-oriented software metrics. IEEE Transactions on Software Engineering
  24(6),  491--496 (1998)

\bibitem{hoyle2001iso}
Hoyle, D.: Iso 9000: quality systems handbook  (2001)

\bibitem{eee1990standard}
IEEE: Standard glossary of softwareengineering terminology. IEEE Software
  Engineering Standards Collection. IEEE pp. 610--12 (1990)

\bibitem{ISO25010}
{ISO/IEC}: {ISO/IEC 25010 - Systems and software engineering - Systems and
  software Quality Requirements and Evaluation (SQuaRE) - System and software
  quality models}. Tech. rep. (2010)

\bibitem{li1993object}
Li, W., Henry, S.: Object-oriented metrics that predict maintainability.
  Journal of systems and software  23(2),  111--122 (1993)

\bibitem{marchesi1998ooa}
Marchesi, M.: Ooa metrics for the unified modeling language. In: Software
  Maintenance and Reengineering, 1998. Proceedings of the Second Euromicro
  Conference on. pp. 67--73. IEEE (1998)

\bibitem{mccabe1976complexity}
McCabe, T.J.: A complexity measure. IEEE Transactions on software Engineering
  (4),  308--320 (1976)

\bibitem{meseguer2006specification}
Meseguer, J., Sharykin, R.: Specification and analysis of distributed
  object-based stochastic hybrid systems. In: International Workshop on Hybrid
  Systems: Computation and Control. pp. 460--475. Springer (2006)

\bibitem{michel2009multi}
Michel, F., Ferber, J., Drogoul, A., et~al.: Multi-agent systems and
  simulation: a survey from the agents community’s perspective. Multi-Agent
  Systems: Simulation and Applications, Computational Analysis, Synthesis, and
  Design of Dynamic Systems pp. 3--52 (2009)

\bibitem{moody2005theoretical}
Moody, D.L.: Theoretical and practical issues in evaluating the quality of
  conceptual models: current state and future directions. Data \& Knowledge
  Engineering  55(3),  243--276 (2005)

\bibitem{mylopoulos1992conceptual}
Mylopoulos, J.: Conceptual modelling and telos. Conceptual Modelling,
  Databases, and CASE: an Integrated View of Information System Development,
  New York: John Wiley \& Sons pp. 49--68 (1992)

\bibitem{nelson2012conceptual}
Nelson, H.J., Poels, G., Genero, M., Piattini, M.: A conceptual modeling
  quality framework. Software Quality Journal  20(1),  201--228 (2012)

\bibitem{oman1992metrics}
Oman, P., Hagemeister, J.: Metrics for assessing a software system's
  maintainability. In: Software Maintenance, 1992. Proceerdings., Conference
  on. pp. 337--344. IEEE (1992)

\bibitem{riaz2009systematic}
Riaz, M., Mendes, E., Tempero, E.: A systematic review of software
  maintainability prediction and metrics. In: Proceedings of the 2009 3rd
  International Symposium on Empirical Software Engineering and Measurement.
  pp. 367--377. IEEE Computer Society (2009)

\bibitem{rizvi2010maintainability}
Rizvi, S., Khan, R.: Maintainability estimation model for object-oriented
  software in design phase (memood). arXiv preprint arXiv:1004.4447  (2010)

\bibitem{wolfgang1994design}
Wolfgang, P.: Design patterns for object-oriented software development.
  Reading, Mass.: Addison-Wesley (1994)

\end{thebibliography}

\end{document}